\documentclass[sigplan,10pt,nonacm]{acmart}
\setcopyright{none}

\usepackage{booktabs}   
\usepackage{subcaption} 
\usepackage[percent]{overpic}
\usepackage{xspace}
\usepackage{amsthm}
\usepackage[linesnumbered]{algorithm2e}
\usepackage{empheq}
\usepackage{amsmath}
\usepackage{pifont}
\usepackage{array}
\usepackage{multirow}

\newtheorem{lemma}{Lemma}

\newcommand{\rlibm}{\textsc{RLibm}\xspace}

\newcommand{\tool}{\textsc{RLibm-Prog}\xspace}

\newcommand{\eg}{\emph{e.g.}\xspace}
\newcommand{\ie}{\emph{i.e.}\xspace}

\newcommand{\rlibmall}{\textsc{RLibm-All}\xspace}

\newcommand{\RNE}{\textit{rn}\xspace}
\newcommand{\RNA}{\textit{ra}\xspace}
\newcommand{\RNZ}{\textit{rz}\xspace}
\newcommand{\RNP}{\textit{ru}\xspace}
\newcommand{\RNN}{\textit{rd}\xspace}

\newcommand{\cmark}{\ding{51}}
\newcommand{\xmark}{\textcolor{red}{\ding{55}}}
\newcommand{\NA}{N/A}
\newcolumntype{P}[1]{>{\centering\arraybackslash}p{#1}}
\newcommand{\bfloat}{BF16\xspace}
\newcommand{\tensorfloat}{TF32\xspace}
\newcommand{\float}{FP32\xspace}

\begin{document}

\title[]{\tool: Progressive Polynomial Approximations for Fast Correctly
  Rounded Math Libraries\\ \small{Rutgers Department of Computer Science Technical Report DCS-TR-758}}

\author{Mridul Aanjaneya}
\orcid{0000-0002-5286-8173}             
\affiliation{
  \department{Department of Computer Science}              
  \institution{Rutgers University}            
  \country{United States}                    
}
\email{mridul.aanjaneya@rutgers.edu}          

\author{Jay P. Lim}
\orcid{0000-0002-7572-4017}             
\affiliation{
  \department{Department of Computer Science}              
  \institution{Yale University}            
  \country{United States}                    
}
\email{jay.lim@yale.edu}          

\author{Santosh Nagarakatte}
\orcid{0000-0002-5048-8548}             
\affiliation{
  \department{Department of Computer Science}              
  \institution{Rutgers University}            
  \country{United States}                    
}
\email{santosh.nagarakatte@cs.rutgers.edu}          

\begin{abstract}
This paper presents a novel method for generating a single polynomial
approximation that produces correctly rounded results for all inputs
of an elementary function for multiple representations. The generated
polynomial approximation has the nice property that the first few
lower degree terms produce correctly rounded results for specific
representations of smaller bitwidths, which we call \emph{progressive}
performance. To generate such progressive polynomial approximations,
we approximate the correctly rounded result and formulate the
computation of correctly rounded polynomial approximations as a linear
program similar to our prior work on the \rlibm project.
To enable the use of resulting polynomial approximations in mainstream
libraries, we want to avoid piecewise polynomials with large lookup
tables.
We observe that the problem of computing polynomial approximations for
elementary functions is a linear programming problem in low
dimensions, \ie, with a small number of unknowns.
We design a fast randomized algorithm for computing polynomial
approximations with progressive performance. Our method produces
correct and fast polynomials that require a small amount of storage. A
few polynomial approximations from our prototype have already been
incorporated into LLVM's math library.
\end{abstract}

\maketitle

\section{Introduction}

Correct rounding of primitive arithmetic operations is mandatory for
floating-point (FP) implementations since the inception of the IEEE
754 standard.  This requirement was not enforced for elementary
functions (algebraic functions such as $1/\sqrt{x}$ and transcendental
functions such as $\sin$, $\cos$, $\log$, $\exp$, etc.)  due to the
\emph{Table Maker's Dilemma}~\cite{Muller:elemfunc:book:2005}.  When
the output of an elementary function matches the result that is
computed with infinite precision and rounded to the target
representation, then it is a correctly rounded result. Correctly
rounded elementary functions can enhance the reproducibility and
portability of software systems.  The IEEE 754-2008 standard has
recommended (yet not mandated) correct rounding of elementary
functions.  Research efforts from several groups have shown that
correctly rounded elementary functions can be obtained at a
``reasonable''
cost~\cite{Lefevre:toward:tc:1998,Daramy:crlibm:spie:2003,Daramy:crlibm:doc}.
Yet, mainstream math libraries for a 32-bit float still do not produce
correctly rounded results for all inputs. When correctly rounded
libraries for double precision such as CR-LIBM are re-purposed for
32-bit floats, they can produce wrong results due to double rounding
errors.

\begin{figure*}[t]%
	\includegraphics[width= 0.8\textwidth]{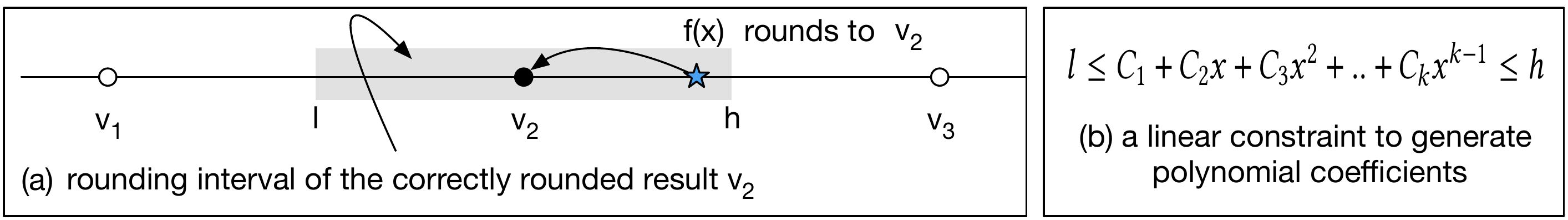}
	\caption{Illustration of the \rlibm approach. (a) The values
          $v_1$, $v_2$, and $v_3$ are representable in the FP
          representation $\mathbb{T}$. The real value of $f(x)$ for a
          given input $x$ cannot be exactly represented in
          $\mathbb{T}$ and is rounded to $v_2$. The \rlibm approach
          identifies the rounding interval of $v_2$ and computes polynomial approximations that produce values in
          this interval. (b) A linear constraint that the polynomial approximation with $k$ terms
          must satisfy for each rounding interval (\ie, $[l, h]$)
          for each input $x$.}
	\label{fig:rlibm_explain}
\end{figure*}

We have been building correctly rounded functions as part of the
\rlibm project~\cite{rlibm-project,
  lim:rlibm32:pldi:2021,lim:rlibm:popl:2021,lim:rlibmall:popl:2022,
  Lim:rlibm:arxiv:2020,Lim:rlibm32:arxiv:2021,Lim:rlibmAll:arxiv:2021,lim:rlibm:phdthesis:2021}. Our
key insight in the \rlibm project is to separate the task of
generating the oracle of an elementary function from the task of
generating efficient implementations. Given an oracle (\eg, a high
precision math library), we make a case for approximating the
correctly rounded result rather than the real value of an elementary
function to generate efficient
implementations. Figure~\ref{fig:rlibm_explain}(a) shows the real
value and the correctly rounded result for an input.  There is an
interval of real values around the correctly rounded result such that
all real values round to it, which is called the \emph{rounding
interval}. This interval provides the constraints on the result of the
polynomial approximation for a given input (see
Figure~\ref{fig:rlibm_explain}(b)).  Next, we formulate the task of
identifying the coefficients of a polynomial of a specific degree that
produces a value in the rounding interval for all inputs as a system
of linear inequalities.

By approximating the correctly rounded result, \rlibm provides more
freedom, allows for lower degree polynomial approximations, and can be
realized via a carefully-crafted system of linear inequalities. One
challenge with the \rlibm approach is that modern LP solvers can only
handle a few thousand constraints. Hence, our \rlibm prototypes create
piecewise polynomials for $32$-bit types. Such piecewise polynomials
are created for each function and for each representation and rounding
mode. We have shown that the resulting functions are both correctly
rounded and faster than mainstream libraries such as Intel's libm and
glibc's libm~\cite{lim:rlibm32:pldi:2021}.

A recent result from our \rlibm project,
\rlibmall~\cite{lim:rlibm:phdthesis:2021,lim:rlibmall:popl:2022},
generates a single polynomial approximation that produces correctly
rounded results for multiple representations and rounding modes.  The
key idea behind \rlibmall is to generate a polynomial approximation
that produces correctly rounded results for a floating-point (FP)
representation with two additional bits of precision (\ie, $n+2$-bits)
using the \emph{round-to-odd} mode. The resulting polynomial
approximation produces correctly rounded results for all five rounding
modes in the standard and for multiple representations with $k$-bits
of precision where $\vert E\vert+1<k\leq n$ and $\vert E\vert$ is the
number of exponent bits in the representation.
The \rlibmall prototype also generates piecewise
polynomials~\cite{Lim:rlibmAll:arxiv:2021,lim:rlibmall:popl:2022}.
A single generic polynomial approximation that produces correct
results for multiple representations and rounding modes is attractive
because it avoids unnecessary code duplication and can enable adoption
by mainstream libraries.

Although polynomial approximations resulting from various \rlibm
prototypes are fast and correct, they had not been incorporated into
mainstream libraries because of the large lookup tables required for
the piecewise polynomials. Space usage by the mainstream library is an
important consideration as these libraries are used in numerous
domains ranging from micro-controllers to high performance systems. To
enable mainstream usage of polynomial approximations from the \rlibm
project, we want to avoid generating large piecewise polynomials and
the accompanying lookup tables. Further, we want to improve
performance for representations with fewer bits rather than every
representation having the same performance because low bitwidth
representations are increasingly used in various
domains~\cite{Tagliavini:bfloat:date:2018,
  nvidia:tensorfloat:online:2020}.

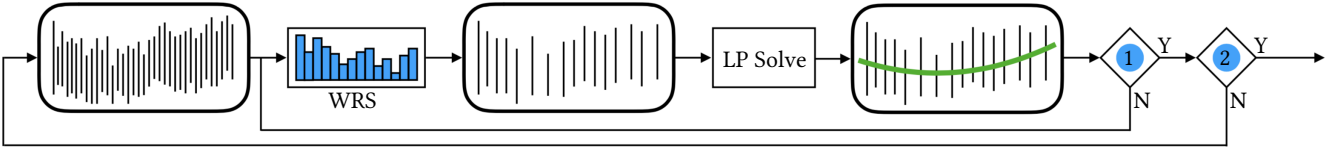
\begin{figure*}[t!]
\vspace*{-2mm}
\begin{overpic}[width=\textwidth]{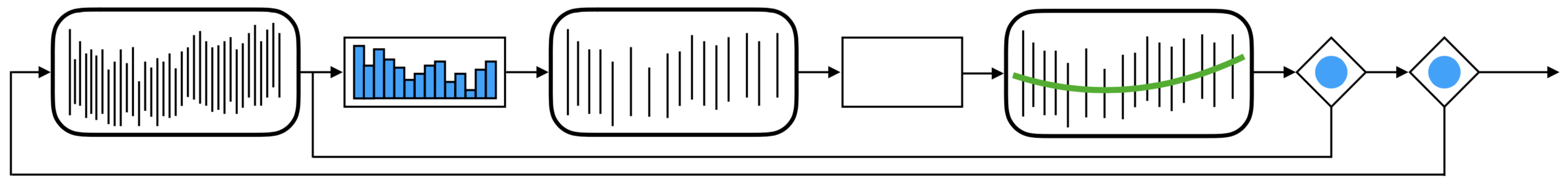}
    \put(54.5,6.6){\small LP Solve}
    \put(25,3.5){\small WRS}
    \put(87,7.6){\small Y}
    \put(94.2,7.6){\small Y}
    \put(84.4,6.7){\small 1}
    \put(91.6,6.7){\small 2}
    \put(85.2,3.5){\small N}
    \put(92.4,3.5){\small N}
\end{overpic}
\vspace*{-8mm}
\caption{\small Overview of our workflow. From the initial list of
  intervals, we perform a weighted random sampling (WRS) to compute a
  smaller subset of intervals, which is then passed to a
  high-precision LP solver. If the LP solution violates more than
  $1/3k$ of the total set of intervals, then we discard the solution
  and perform WRS again (loop 1). Otherwise, we check if the LP
  solution has $0$ or a small number of violated intervals. If not, we
  double the weights for the violated intervals and repeat the whole
  process for a fixed number of iterations (which we show is guaranteed to
  converge if the system of constraints is full-rank) before increasing the
  number of terms in the polynomial (loop 2).}
\label{fig:workflow}
\end{figure*}

\textbf{Progressive polynomials.} Our goal is to generate a single polynomial
approximation that produces correctly rounded results for multiple
FP representations with \emph{progressive}
performance. The first few lower degree terms of such a polynomial
produces correctly rounded results for representations with fewer
precision bits and the higher degree terms become necessary for
representations with more precision bits, while keeping the
\emph{same} lower degree terms. We call such polynomials \emph{progressive
polynomials}. They are inspired by Taylor
polynomials, which provide better polynomial fits as one uses more
terms. These progressive polynomial approximations offer two major
benefits. First, they will provide more efficient implementations for
representations with fewer precision bits (\eg,
bfloat16~\cite{Tagliavini:bfloat:date:2018} or
tensorfloat32~\cite{nvidia:tensorfloat:online:2020}) in comparison to
\rlibmall that uses the same high degree polynomial approximation
across all representations (see Section 4). Second, they will provide a unified
approach to implementing math library functions, as representations
with less precision bits can \emph{reuse} the implementation for those
with more precision bits, while discarding the higher order terms from
the polynomial.

For example, consider the case when we want to produce a single
approximation for $e^x$ that produces correct results for all inputs
in the 32-bit float, 16-bit bfloat16, and 19-bit tensorfloat32
types. For the sake of argument, suppose we generate a 6-term,
$5^{th}$-degree progressive polynomial ($C_{1} + C_{2}x^{1} +
C_{3}x^{2} + C_4 x^3 + C_5 x^4 + C_6 x^5$). We use all 6-terms of the
polynomial to produce correctly rounded results for a 32-bit float
input. We use only the first four terms of the polynomial to produce
correct results for a tensorfloat32 input, which is faster than
producing the result for a 32-bit float.  Similarly, we use only the
first three terms to generate correctly rounded results for a bfloat16
input, which is faster than producing results for both tensorfloat32
and float inputs.

\textbf{Efficient randomized algorithm for solving linear constraints.} To
generate progressive polynomial approximations and to avoid storing
large tables of coefficients for piecewise polynomials, we observe
that the problem of computing a polynomial approximation using the
linear programming approach of \rlibm is a linear program in \emph{low
dimensions}, with far fewer unknown
variables in comparison to the number of constraints.  Inspired by
prior work on linear programs in low
dimensions~\cite{clarkson:vegas:jacm:1995}, we design a fast
randomized algorithm for producing progressive polynomial
approximations that uses a significantly smaller table of coefficients
(by an order of magnitude) compared to \rlibmall.

Given the number of terms for each representation of interest used in
the progressive polynomial, our algorithm uses an LP solver to only
solve a small set of $6k^2$ constraints, where $k$ is the maximum
number of terms used in the progressive polynomial. Given a multi-set
of constraints, the algorithm samples $6k^2$ constraints from the
entire set of constraints and solves the sample optimally using the LP
solver. If the sample solution violates more than $1/3k$ of the
multi-set, it discards the sample. If the sample solution violates
less than $1/3k$ of the multi-set, it adds the violated constraints
once more to the sample. To efficiently implement this algorithm, we
use weights to encode the multi-set and use weighted random sampling
to create the sample~(see Section~\ref{approach:polygen}).

This process repeats until we find a solution that satisfies all
constraints, which happens when the system of linear inequalities is
full-rank (\ie, there are $k$-linearly independent constraints) or
when the number of iterations reaches a threshold. Since we do not
know the rank of our system, we iteratively increase the number of
terms used for the polynomial and have a threshold on the number of
iterations. When the system is full-rank, we prove that our algorithm
finds the progressive polynomial in $6k\log n$ iterations in
expectation (see Section~\ref{proof}). Figure~\ref{fig:workflow}
illustrates the process of solving the set of linear inequalities with
our approach.

\textbf{Prototype and results.} Our prototype, \tool, provides a
single progressive polynomial approximation that produces the
correctly rounded results for multiple representations and multiple
rounding modes for 10 elementary functions. It has progressive
performance with bfloat16 and tensorfloat32 inputs being 25\% and 16\%
faster than evaluating the entire polynomial.
The randomized algorithm produces polynomial approximations that
require an order of magnitude lower storage than prior \rlibm
prototypes~\cite{lim:rlibmall:popl:2022,lim:rlibm32:pldi:2021}.
\tool's polynomials for the 32-bit float type are faster than all
mainstream and/or correctly rounded libraries.
Three polynomial approximations ($ln(x)$, $log_2(x)$, and
$log_{10}(x)$) generated by our prototype have already been
incorporated into LLVM's math
library~\cite{ly:logf:rlibm:2021,ly:log2f:rlibm:2021,ly:log10f:rlibm:2021}.

\section{Background}
\begin{figure*}[t!]
    \centerline{\includegraphics[width=1.0\linewidth]{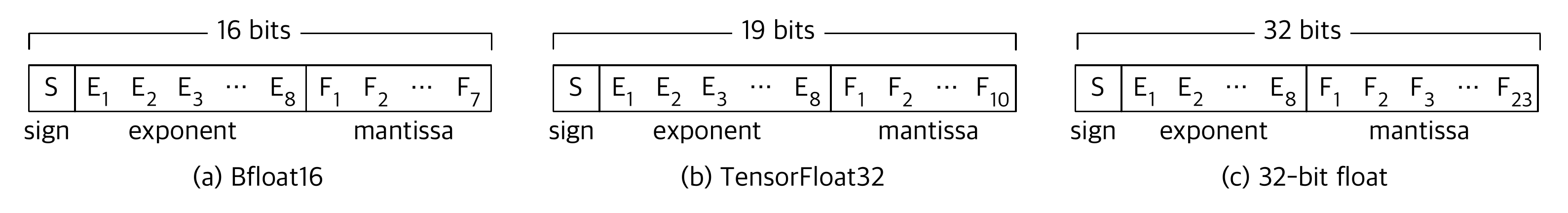}}
  \caption{\small The three FP representations used in this paper: (a)
    Bfloat16, (b) TensorFloat32, and (c) 32-bit float.}
  \label{fig:fp_conf_rep}
\end{figure*}

We provide background on the FP
representation, the process of computing polynomial approximations for
elementary functions, and the \rlibm
approach~\cite{lim:rlibm32:pldi:2021,lim:rlibm:popl:2021,lim:rlibmall:popl:2022,
  Lim:rlibm:arxiv:2020,Lim:rlibm32:arxiv:2021,Lim:rlibmAll:arxiv:2021,lim:rlibm:phdthesis:2021}.

\subsection{The Floating-Point Representation}
\label{sec:finiteprecision}
The IEEE-754 standard specifies the FP representation $\mathbb{F}_{n,
  |E|}$ that is parameterized based on the total number of bits ($n$)
and the number of bits used for the exponent ($|E|$). The goal is to
represent a large range of values (\ie, wider dynamic range) with
a reasonable amount of accuracy (\ie, precision)~\cite{goldberg:fp}. The
sign of a value is specified by a dedicated sign bit ($S$). To
represent a large range of values, the FP representation has an
unsigned exponent field ($E$). Each value is represented as precisely
as possible with the mantissa bits ($F$).
Figure~\ref{fig:fp_conf_rep}(c) depicts the bit-string for a 32-bit
float.

The values represented by the FP representation are classified into
three classes: (a) normal values when the exponent field is neither
all zeros nor all ones, (b) subnormal or denormal values when the
exponent field is all zeros, and (c) special values when the exponent
field is all ones. In the case of normal values, the value represented
by the FP bit-string is $(1 + \frac{F}{2^{|F|}})\times 2^{E-bias}$,
where $bias$ is $2^{|E| - 1} - 1$. With subnormal values, the value
represented by the bit-string is $(\frac{F}{2^{|F|}})\times
2^{1-bias}$. Subnormal values are used to represent values close to
zero. In the case of special values, when the mantissa bits are all
zeros, then the bit-string represents positive or negative infinity
depending on the sign bit. Otherwise, the bit-string represents
not-a-number (NaN).

The common formats specified in the IEEE-754 standard are 16-bit half
precision ($\mathbb{F}_{16, 5}$), 32-bit single precision float
($\mathbb{F}_{32, 8}$), and 64-bit double precision ($\mathbb{F}_{64,
  11}$).

\textbf{The bfloat16 and tensorfloat32 formats.}  Numerous recent
variants of the IEEE-754 FP representation increase either the dynamic
range or the precision when compared to the existing half precision
format. The bfloat16 format~\cite{Tagliavini:bfloat:date:2018} is a
16-bit representation with 8 bits for the exponent (\ie,
$\mathbb{F}_{16, 8}$).
Nvidia's \texttt{tensorfloat32}~\cite{nvidia:tensorfloat:online:2020}
is a 19-bit representation with 8-bits for the exponent~(\ie,
$\mathbb{F}_{19, 8}$). It provides the dynamic range of bfloat16 and
the precision of the half precision format.
Figure~\ref{fig:fp_conf_rep}(a) and Figure~\ref{fig:fp_conf_rep}(b)
show the bfloat16 and the tensorfloat32 format.

\textbf{Rounding mode.} When a real value is not exactly representable
in the FP representation, it needs to be rounded to a value in the FP
representation. The IEEE-754 standard specifies five distinct rounding
modes that rounds the real value to one of the two adjacent FP values:
round-to-nearest-ties-to-even (\RNE), round-to-nearest-ties-to-away
(\RNA), round-towards-zero (\RNZ), round-towards-positive-infinity
(\RNP), and round-towards-negative-infinity (\RNN). Different rounding
modes have different trade-offs in the implementation of various FP
operations. The \RNE mode is the widely used rounding mode.

\subsection{Approximating Elementary Functions}
Elementary functions are functions of a single variable that are
typically approximated with polynomial approximations. It is feasible
to design polynomial approximations with low error for an elementary
function when the input domain is small.  Hence, one of the crucial
steps in approximating any elementary function is range reduction.

\textbf{Range reduction and output compensation.} Range reduction
reduces the domain of an elementary function $f(x)$ to a small input
domain using mathematical identities~\cite{Cody:book:1980}. The range
reduction transforms an input $x$ from the original domain of inputs
to a reduced input $x'$. The polynomial approximations are performed
with the reduced inputs~(\ie, $y' = P(x')$). After range reduction, the
function being approximated with polynomial approximation may be
different from the original elementary function (\eg, $ln(x)$ can be
approximated with $log_2(x')$). 
The output ($y'$) has to be adjusted appropriately to produce the
output for the original input ($x$). The output compensation function
produces the final result by compensating the range reduced output
$y'$ based on the range reduction performed for input $x$. 

\textbf{Polynomial approximations.} The next step is to generate
polynomial approximations that take reduced inputs and produce the
result of the elementary function in the reduced input domain.  One
common method to generate such polynomial approximations is to
minimize the maximum error of the polynomial approximation with
respect to the real value of the elementary function (also known as
minimax approximations~\cite{Muller:elemfunc:book:2005}). A commonly
used mini-max approximation is the Remez
algorithm~\cite{Remes:algorithm:1934}. Using real analysis, one can
bound the maximum error of such a minimax approximation.
CR-LIBM~\cite{Daramy:crlibm:spie:2003,Daramy:crlibm:doc}, a correctly
rounded library for the double precision type, uses this near-minimax
approach to generate polynomial approximations.

Range reduction, output compensation, and polynomial evaluation are
all implemented in a finite precision representation. Hence, they can
experience numerical errors, which when coupled with polynomial
approximation errors can cause wrong results.

\subsection{The \rlibm Approach}
We provide a brief background on our prior work in the \rlibm
project~\cite{lim:rlibm32:pldi:2021,lim:rlibm:popl:2021,lim:rlibmall:popl:2022,
  Lim:rlibm:arxiv:2020,Lim:rlibm32:arxiv:2021,Lim:rlibmAll:arxiv:2021,lim:rlibm:phdthesis:2021},
where we decouple the problem of generating an oracle from the task of
the generating efficient implementations.  We assume the existence of
an oracle~(which may be slow) that provides correctly rounded results.
This oracle is only used to compute the correctly rounded result of an
elementary function $f(x)$ for each input $x$ in the target
representation $\mathbb{T}$.
Once there is an oracle result, the \rlibm project makes a case for
approximating the correctly rounded result rather than the real value
of an elementary
function~\cite{lim:rlibm32:pldi:2021,lim:rlibm:popl:2021}.
%
%
An FP representation can only represent finitely many values
accurately. Hence, there is an interval of real values around the
correctly rounded result such that all values in the interval round to
it.  This is the maximum amount of freedom available for the
polynomial approximation.  The \rlibm project has demonstrated that
this amount of freedom for polynomial generation by approximating the
correctly rounded result is much larger than the one with the minimax
approach. Hence, \rlibm prototypes provide significant performance
benefits when compared to highly optimized
libraries~\cite{lim:rlibm32:pldi:2021}.

Given the correctly rounded result, the next step is to identify an
interval $[l, h]$ around the correctly rounded result such that any
value in $[l, h]$ rounds to the correctly rounded result, which is
called the \textit{rounding interval}.
Figure~\ref{fig:rlibm_explain}(a) illustrates the rounding interval
around the correctly rounded result.
Next, range reduction specific to the elementary function is applied
to transform an input $x$ to $x'$. The polynomial approximation will
approximate the result for $x'$. To perform polynomial approximation,
one needs the rounding interval that corresponds to the reduced input
$x'$. The \rlibm project uses the inverse of the output compensation
function to identify the reduced interval $[l', h']$.

Once a set of reduced intervals is available, the next task is to
synthesize the coefficients of the polynomial with $k$ terms using an
arbitrary precision linear programming (LP) solver such that it
satisfies the reduced constraints (\ie, $l' \leq P(x') \leq h'$).
Figure~\ref{fig:rlibm_explain}(b) shows the linear constraint to
generate the coefficients of a polynomial with $k$ terms.

Subsequently, the result for the original input $x$ is computed with
output compensation.  Range reduction, output compensation, and the
polynomial evaluation happen in some finite precision representation
(\eg, double) and can experience numerical errors. The rounding
intervals are further constrained to ensure that the resulting polynomial
always produces the correctly rounded results for all inputs.

\begin{figure}[!t]%
	\centering
	\includegraphics[width=\linewidth]{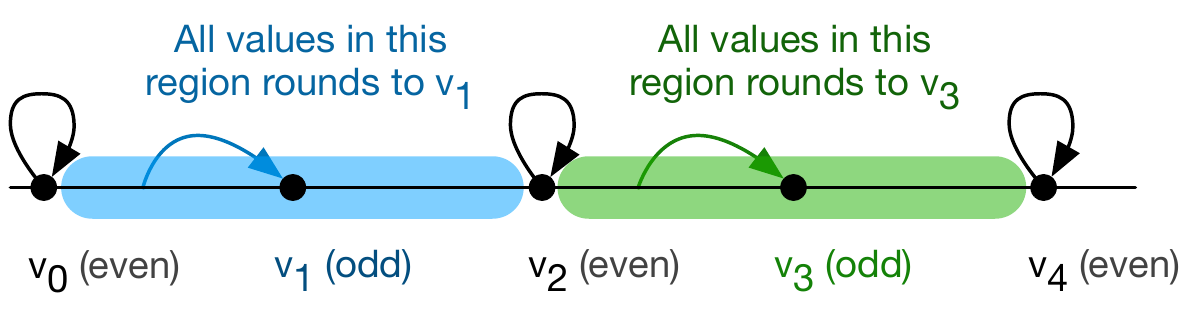}
	\caption{We show the rounding of a real value with the
          round-to-odd mode.  Here, $v_0$, $v_1$, $v_2$, $v_3$, and
          $v_4$ are values representable in a representation
          $\mathbb{T}$. If the real value is exactly representable in
          $\mathbb{T}$, then it rounds to that value. Otherwise, it
          rounds to the nearest value in $\mathbb{T}$ that is odd.}
	\label{fig:rno_explain}
\end{figure}

\textbf{\rlibmall.} The approach described above produces correctly
rounded results for all inputs for a specific rounding mode and
representation.
Our recent work, \rlibmall~\cite{lim:rlibmall:popl:2022}, generates a
single polynomial approximation that produces correctly rounded
results for multiple representations and multiple rounding modes.
When the goal is to create correctly rounded results for a
representation with $n$-bits, the key idea behind \rlibmall is to
create polynomial approximations that produce the correctly rounded
result of $f(x)$ with the \emph{round-to-odd} mode for a
representation with $n+2$-bits (\ie, two additional bits of precision
with the same exponent). We have proven that the resulting polynomial
produces correctly rounded results for all rounding modes in the
standard and all representations with $k$-bits such that $|E| + 1 < k
\leq n$, where $|E|$ is the number of exponent bits. The round-to-odd
mode is a non-standard rounding mode that avoids double rounding
errors and can be described as follows. If the real value is exactly
representable in the target representation, then it is rounded to that
value. Otherwise, it is rounded to an adjacent value whose bit-string
is odd when interpreted as an unsigned
integer. Figure~\ref{fig:rno_explain} pictorially depicts the
round-to-odd mode. To correctly round any real value to a FP
representation with the standard rounding modes, one needs to identify
if the real value is less than, greater than, or equal to the midpoint
of two adjacent FP values. The round-to-odd mode preserves this
information and avoids double rounding
errors~\cite{lim:rlibmall:popl:2022}.

One drawback of the single polynomial approximation with the
round-to-odd mode in \rlibmall is that \emph{every} representation
must pay the computational cost of the largest representation. Our
\rlibmall prototype generated piecewise polynomials with large lookup
tables because we were not aware of an effective method to solve a
large number of constraints at that point in time and LP solvers
cannot automatically solve millions of constraints.  These piecewise
polynomials require large lookup tables of coefficients that may not
be ideal in resource-constrained environments.

\section{Progressive Polynomial Approximations}
Our goal is to generate a single polynomial approximation that not
only produces correctly rounded results for multiple representations
and rounding modes but also has progressively better performance for
lower bitwidth representations given a set of representations. We call
them progressive polynomials.  If we can generate such progressive
polynomials, then we can evaluate the first few terms of the
polynomial to obtain the correct results for lower bitwidth
representations and the entire polynomial for the largest
representation.

This paper proposes a novel method to generate progressive polynomial
approximations.
Building on our prior work in the \rlibm
project~\cite{lim:rlibm32:pldi:2021,lim:rlibm:popl:2021,lim:rlibmall:popl:2022,
  Lim:rlibm:arxiv:2020,Lim:rlibm32:arxiv:2021,Lim:rlibmAll:arxiv:2021,lim:rlibm:phdthesis:2021},
we approximate the correctly rounded result and use a linear
programming formulation to generate polynomial approximations.
In contrast to our prior work in the \rlibm project, our setting has
significantly larger number of constraints (a constraint for each
input and for each type) because we are generating progressive
polynomials.
In our prior work in the \rlibm project, we were not aware of an
effective way to solve an LP problem with millions of
constraints. Hence, our prior \rlibm prototypes generated piecewise
polynomials with large lookup tables to store the polynomial
coefficients.  The presence of these lookup tables was a barrier for
adoption of our polynomial approximations into mainstream math
libraries. Hence, we do not want to generate large piecewise
polynomials.

A key observation that we make in this paper is that the system of
linear inequalities generated by the \rlibm approach is a linear
program in low dimensions (\ie, a polynomial with a small number of
terms $k$ that satisfies millions of constraints).
If the set of linear constraints is full-rank, then there exist $k$
linearly independent constraints that identify the polynomial
coefficients~\cite{clarkson:vegas:jacm:1995}.
Our goal is to develop a fast iterative method for generating
progressive polynomials without large lookup tables. One challenge in
this setting is that we do not know the rank $k$ of the set of
constraints.

\subsection{Overview of Our Method}
Our approach for generating progressive polynomial approximations
consists of the following steps. First, we iteratively explore the
number of terms for each individual representation of interest in our
progressive polynomial. Second, we use an oracle (\ie, an existing
high-precision library) to identify the correctly rounded result for
each representation.  For the largest representation $\mathbb{T}_{i}$
of interest, we generate correctly rounded results for a
representation with two additional bits of precision
($\mathbb{T}_{i+2}$) with the \emph{round-to-odd} mode inspired by our
prior work on \rlibmall~\cite{lim:rlibmall:popl:2022}. The resulting
polynomial approximation produces correctly rounded results for all
representations $\mathbb{T}_{j}$, where $j \leq i$, and for all
rounding modes as long as $\mathbb{T}_{j}$ has the same number of
exponent bits as $\mathbb{T}_i$.

Third, we identify an interval of real values that round to the
correctly rounded result for every input, which is known as the
rounding interval. Fourth, we perform range reduction to identify the
reduced input and infer the reduced rounding intervals. Subsequently,
we attempt to generate a progressive polynomial from the set of
reduced inputs and reduced rounding intervals for each
representation. We generate constraints for the largest representation
that uses all terms of the polynomial. The polynomial when evaluated
should produce a value in the reduced rounding interval. For other
representations, we systematically hypothesize a specific number of
terms for generating the progressive polynomial.
Fifth, we try to generate a polynomial that has $k$~terms and is of
degree~$d$ given $n$ constraints (\eg, $n$ is 512 million with
$e^x$).
We extend Clarkson's method~\cite{clarkson:vegas:jacm:1995} to our
context and develop a fast randomized algorithm to identify $k$
linearly independent constraints that identifies the polynomial.
If the systems of linear inequalities has full rank, then it has a
unique solution.

Our randomized algorithm can be described as follows. Initially, we
maintain a multiset $M$ of all $n$ constraints. We sample $6k^2$
constraints from $M$, where $k$ is the total number of terms for the
largest representation in the progressive polynomial. We solve the
sample optimally using an LP solver to obtain the solution $x^*$.
The LP solver solves the sample with real values but the eventual
polynomial evaluation happens in double precision. We check if the
sample solution $x^*$ satisfies all inputs in the sample when
evaluated in double precision. If not, we restrict the rounding
interval for the input that is not satisfied by $x^*$. We subsequently
attempt to solve the sample with the revised constraint.
If we are able to solve the sample, then we use the resulting
polynomial to identify constraints in $M$ that are not satisfied by
$x^*$.  While checking whether $x^*$ satisfies the constraint, we
evaluate only the specified number of terms as dictated by the
configuration of the progressive polynomial.  If more than $1/3k$ of
the set $M$ of constraints is not satisfied by $x^*$, then we discard
the sample and repeat the above process by creating a new
sample. Otherwise, we add each constraint that was not satisfied one
additional time to the multiset $M$. We repeat the above process until
we find that $x^*$ for the sample does not violate any constraint in
$M$ or the number of iterations exceeds the user-specified cut-off. If
there exists a solution (\ie, the system of linear inequalities is
full-rank), then the above algorithm finds it in $6k\log(n)$
iterations in expectation.

Our procedure for computing the oracle result, identifying the
rounding intervals, and deducing the reduced rounding intervals is
identical to our prior work in the \rlibm
project~\cite{lim:rlibm32:pldi:2021,lim:rlibm:popl:2021,lim:rlibmall:popl:2022}.
The key difference lies in the manner in which we generate linear
constraints for progressive polynomials, the manner in which we
evaluate polynomials, and our procedure for generating the polynomial
approximation given a set of linear constraints.

\subsection{Linear Constraints for Progressive Polynomials}
A reduced input $x$ can be present in multiple representations. The
rounding interval for each such reduced input will be different
depending on the representation(\ie, $[l^{\mathbb{T}_1}_{x},
  h^{\mathbb{T}_1}_{x}]$ for representation $\mathbb{T}_1$ and
$[l^{\mathbb{T}_2}_{x}, h^{\mathbb{T}_2}_{x}]$ for representation
$\mathbb{T}_2$). A representation with lower bitwidths will have
larger rounding intervals as the spacing between adjacent points is
relatively larger when compared to a representation with larger
bitwidth. We want a single polynomial approximation to satisfy all
these bounds of the rounding intervals. Hence,

\begin{small}
\begin{align}
  \nonumber l^{\mathbb{T}_1}_{x} \leq P(x) \leq h^{\mathbb{T}_1}_{x} \\
  \nonumber l^{\mathbb{T}_2}_{x} \leq P(x) \leq h^{\mathbb{T}_2}_{x} \\
  \nonumber l^{\mathbb{T}_3}_{x} \leq P(x) \leq h^{\mathbb{T}_3}_{x}  
\end{align}
\end{small}

\textbf{Progressive performance.} We want the resulting single
polynomial approximation to have better performance while producing
correctly rounded results for lower bitwidths (\ie, progressive performance) when
compared to evaluating the entire polynomial for larger
bitwidths. Given the number of terms for a representation with a
particular bitwidth and the total number of terms for the entire
polynomial approximation, we create constraints such that evaluating
the first few terms produces a value that lies in the rounding
interval corresponding to that representation. Consider the scenario
where $\mathbb{T}_1$ is the representation with the largest bitwidth.
We are trying to find a polynomial approximation with $k_1$ terms for
it.  We also want to find coefficients such that when we evaluate
inputs belonging to representations $\mathbb{T}_2$ and $\mathbb{T}_3$
with $k_2$ and $k_3$ terms (here $k_1 > k_2 > k_3$), they lie within
their respective rounding intervals.  The system of linear constraints
that we generate for a given input $x$ is as follows,

\begin{eqnarray*}
\nonumber  l^{\mathbb{T}_3}_{x} \leq \underbrace{C_1 + C_2 x + \ldots + C_{k_3} x^{k_3-1}}_{\mathcal{P}_3(x)}  \leq h^{\mathbb{T}_3}_{x} \\
\nonumber  l^{\mathbb{T}_2}_{x} \leq \underbrace{\mathcal P_3(x) + \ldots + C_{k_2} x^{k_2-1}}_{\mathcal P_2(x)}  \leq h^{\mathbb{T}_2}_{x} \\
\nonumber  l^{\mathbb{T}_1}_{x} \leq \underbrace{\mathcal P_2(x) + \ldots +  C_{k_1} x^{k_1 -1}}_{\mathcal P_1(x)} \leq h^{\mathbb{T}_1}_{x}
\end{eqnarray*}

When we generate constraints for representation $\mathbb{T}_2$, we use
the exact same coefficients for the first $k_2$ terms as we did for
representation $\mathbb{T}_1$. Similarly, we use the same coefficients
for the first $k_3$ terms for representation $\mathbb{T}_3$. If we are
able to find such polynomials, the resulting polynomial approximation
not only produces correctly rounded results for all inputs but also has better
performance for representations with lower bitwidth when compared to
evaluating all terms in the polynomial.
Note that this formulation for generating progressive polynomials
creates significantly more constraints (since there is a constraint
for each input and each representation). Hence, an efficient method to
generate polynomial approximations is crucial.

\subsection{A Fast Algorithm for Solving Constraints}
\label{approach:polygen}
\begin{algorithm}[t]
\small
\DontPrintSemicolon
\SetKwFunction{FWRS}{WeightedRandomSample}
\SetKwFunction{FCalcRndInterval}{RoundingInterval}
\SetKwFunction{FCalcReducedInterval}{ReducedIntervals}
\SetKwFunction{FGenPoly}{GenProgPolynomial}
\SetKwFunction{FSolveCheck}{SolveSample}
\SetKwProg{Fn}{Function}{:}{}
\Fn{\FGenPoly{$f$, $X$, $RR_{\mathbb{H}}$, $OC_{\mathbb{H}}$,
    $K$, $N$}}{
  $Y \leftarrow \emptyset$\;
  \tcc{Compute the rounding interval}
  \ForEach{$(x, \mathbb{T})  \in X$} {
    $y \leftarrow RN_{\mathbb{T}}(f(x))$\;
    $[l, h] \leftarrow$ \FCalcRndInterval{$y$, $\mathbb{T}$, $\mathbb{H}$}\;
    $Y \leftarrow (x, [l, h])$\;
  }
  
  $\mathcal{L} \leftarrow $ \FCalcReducedInterval{$Y$, $RR_{\mathbb{H}}$, $OC_{\mathbb{H}}$}\;

  \tcc{initialize the weights}
  \ForEach{$x \in \mathcal{L}$} {
    $x.w \leftarrow 1$\;
    $x.u \leftarrow random(0,1)$\;
  }

  $i \leftarrow 0$\;  
  \While {$i < N$}{   
    $S \leftarrow$ \FWRS{$\mathcal{L}$, $K$}\;
    $(poly, n_v)$ $\leftarrow$ \FSolveCheck{$S$, $\mathcal{L}$, $K$}\;
    $i \leftarrow i + 1$\;  
    \If{$n_v < Limit$}{
      \Return{$(poly, n_v)$}\;
    }
  }  
  \Return{$(\emptyset, 0)$}\;
}
\caption{\small Our procedure to generate progressive polynomials for
  an elementary function $f$ given a set of inputs $X$ with their
  respective representations ($\mathbb{T}$).  Range reduction
  ($RR_{\mathbb{H}}$) and output compensation ($OC_{\mathbb{H}}$) are
  performed in representation $\mathbb{H}$. Here, $K$ is a vector that
  provides the number of terms in the progressive polynomial for each
  representation. The maximum number of iterations is specified by
  $N$. We represent the oracle result obtained by rounding the real
  value of $f(x)$ to representation $\mathbb{T}$ by
  $RN_{\mathbb{T}}(f(x))$.  The function \texttt{RoundingInterval}
  computes the rounding interval. The function
  \texttt{ReducedIntervals} computes the reduced inputs and infers the
  reduced intervals. The function \texttt{WeightedRandomSample}
  identifies the sample with weighted random sampling. The function
  \texttt{SolveSample} solves the sample and updates the weights of
  the constraints not satisfied by the solution to the sample, which
  is described in Algorithm~\ref{alg:solvesample}.  }
\label{alg:main}
\end{algorithm}

\begin{algorithm}[t]
\small
\DontPrintSemicolon
\SetKwFunction{FSolveCheck}{SolveSample}
\SetKwProg{Fn}{Function}{:}{}
\Fn{\FSolveCheck{$S$, $\mathcal{M}$, $K$}}{
    $poly \leftarrow LPSolver(S)$\;
    $(w_v, w_s, n_v) \leftarrow (0,0,0)$\;
 
    $k \leftarrow max\_element(K)$\;  
    \ForEach{$(x, [l, h]) \in \mathcal{M}$} {

      \If{$poly(x, K) \in [l,h]$}{
        \tcc{sum the weights of the satisfied constraints}
        $w_s \leftarrow w_s + x.w$\;
      }
      \Else{
        \tcc{sum the weights of the violated constraints}
        $w_v \leftarrow w_v + x.w $\;
        $n_v \leftarrow n_v + 1$\;
      }    
    }
    
    \tcc{Check if it is a lucky iteration}
    \If{$w_v \leq \frac{1}{3k-1} w_s$}{
      \tcc{Double the weights of violated constraints}
      \ForEach{$(x, [l, h]) \in \mathcal{M}$ and $poly(x, K) \notin [l,h]$}{
        $x.w \leftarrow x.w * 2$\;
      }
    }
    \Return{$(poly, n_v)$}\;
}
\caption{\small Given a sample $S$, the total set of reduced inputs
  and constraints $\mathcal{M}$, and the degrees of the progressive
  polynomials, this function \texttt{SolveSample} uses the LP solver
  to solve the sample, identifies whether the iteration happens to be
  a lucky iteration, and doubles the weights of the violated
  constraints on a lucky iteration.  This function returns the
  progressive polynomial that solves the sample and the number of
  constraints violated in $\mathbb{L}$. Here, $poly(x, K)$ evaluates
  the progressive polynomial using the number of terms specified in
  $K$ for various representations with input $x$.}
\label{alg:solvesample}
\end{algorithm}

To create correctly rounded progressive polynomial approximations, our
objective is to generate polynomials of low degree with a few
terms. Our prior work on the \rlibm
project~\cite{lim:rlibm32:pldi:2021,lim:rlibm:popl:2021,lim:rlibmall:popl:2022}
generates piecewise polynomials with approximately $2^{10}$
sub-domains. Such large tables can interfere with caches in
memory-intensive applications and may not be ideal for resource
constrained environments such as micro-controllers.

We make a key observation that our system of linear constraints is a
linear program of small dimensions~\cite{clarkson:vegas:jacm:1995,
  megiddo:jacm:1984}, which is widely studied. We use ideas from prior
work to our setting where we do not know whether the system of linear
constraints is ``full-rank'' (\ie, if there are at least $k$
\emph{linearly independent} constraints).  Further, there can be
several billion constraints. Hence, we have to design memory-efficient
mechanisms to solve them.

Algorithm~\ref{alg:main} describes our procedure to find a progressive
polynomial. As we do not know the rank of our system of linear
constraints, we iteratively increase the number of terms for the
entire polynomial and for the individual representations.  The
procedure to identify a small set of key constraints is as follows:
\begin{itemize}
\item Let $M$ be a multi-set of constraints.
\item \textbf{Step 1}: Sample $S$ constraints from $M$ uniformly at
  random where $|S| = 6k^2$, where $k$ is the number of terms for the
  largest representation with the progressive polynomial.
\item \textbf{Step 2}: Solve the sample $S$ optimally using an LP
  solver to compute $x^*$.
\item \textbf{Step 3}: Check how many constraints of $M$ are not
  satisfied by $x^*$. If more than $1/3k$ of the constraints
  in $M$ are not satisfied by $x^*$, then discard this sample and go
  to Step 1. Otherwise, add all such constraints not satisfied by
  $x^*$ another time to $M$ (\ie, $M$ will now have repeated constraints
  and is a multi-set). We call these iterations \textbf{lucky}
  (\ie, we are making progress towards our goal of
  identifying the crucial $k$ constraints). Then go back to Step 1.
\item Repeat the above until we find a solution $x^*$ that satisfies
  all constraints in $M$ or the number of iterations exceeds the
  user-specified threshold.
\end{itemize}

When we create a sample $S$ with $6k^2$ constraints from the multi-set
$M$ and compute the optimum solution $x^*$ for $S$, then with
probability at least $1/2$, $x^*$ can only violate $1/3k$ of the
constraints in $M$.  We provide a proof that this algorithm is
effective in finding the key constraints necessary to solve the system
of linear constraints quickly in Section~\ref{proof}.

\textbf{Removing the multi-set requirement.}  As we have billions of
constraints in $M$ to start with, maintaining a multi-set in memory is
challenging. Hence, we logically implement such a multi-set by
maintaining weights with each constraint, which are incremented
instead of duplicating constraints. We next describe our procedure to
find a polynomial with $k$ terms using the weight-based
formulation. Initially, each constraint is present only once in the
multi-set version. Hence, we set the weight of each constraint to
1. Subsequently, we sample constraints with probability proportional
to their weights.

\textbf{Weighted random sampling.} We use weighted random
sampling~\cite{wrs:2006} to produce a sample of size $6k^2$ given a
set $M$ with $n$ weighted constraints.
\begin{enumerate}
\item For each constraint $s_i \in M$ with weight $w_i$, set $u_i =
  random(0, 1)$ and $key_i = u_i^{1/w_i}$.
\item Select $6k^2$ items that have the largest values of keys (\ie,
  $key_i$) as the sample.
\end{enumerate}

Here, $u_1$ and $u_2$ are uniform random variables in (0,1). If $X_1 =
u_1^{1/w_1}$ and $X_2 = u_2^{1/w_2}$, then $P( X_1 \leq X_2) = \frac
{w_1}{w_1 + w_2}$. Hence, selecting the largest $6k^2$ items is
equivalent to sampling according to their weights.

\textbf{Identifying the lucky iteration.} The next task in avoiding
the multi-set representation lies in identifying the lucky
iteration. An invariant with our weight-based representation is that
the sum of the weights of all constraints in $M$ is equal to the
cardinality of the multi-set.

\begin{figure*}[t!]
\vspace*{-2mm}
\begin{overpic}[width=\textwidth]{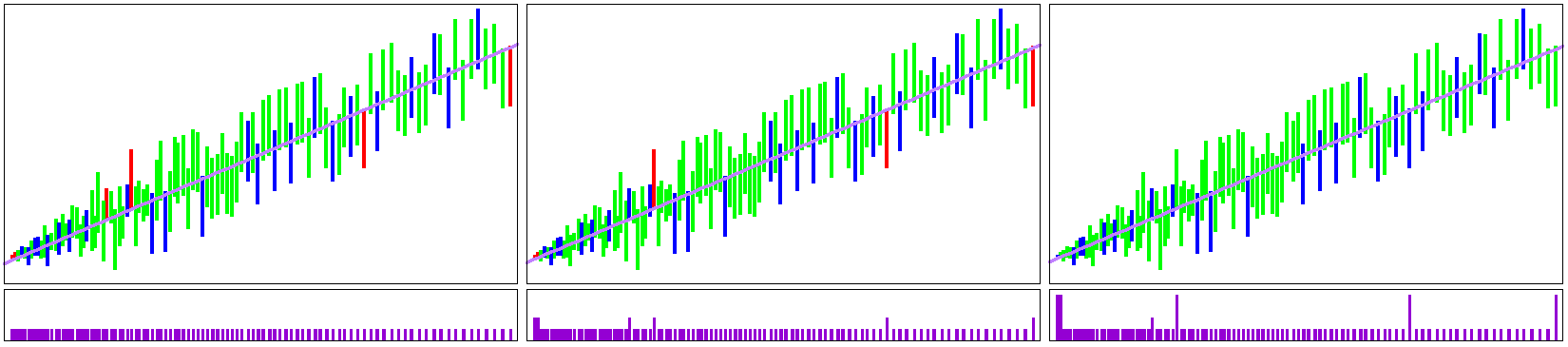}
    \put(1.5,19.5){\small Lucky Iteration: 1}
    \put(34.9,19.5){\small Lucky Iteration: 2}
    \put(68.3,19.5){\small Lucky Iteration: 3}
\end{overpic}
\vspace*{-8mm}
\caption{\small Overview of the weighted random sampling: A linear polynomial with $2$ terms is fit to a set of $100$ intervals.
The $6\cdot2^2=24$ sampled intervals are shown in blue, all satisfied intervals are shown in green, and all violated intervals are shown in red.
Sampled intervals are also satisfied by construction. The computed polynomial is shown in purple, and the weights for each interval are shown as a bar graph.}
\vspace*{-2mm}
\label{fig:wrs}
\end{figure*}

The constraints in $M$ can be divided into two categories: constraints
violated by the sample solution (\ie, $VIO_{x^*}(M)$) and constraints
that are satisfied by the sample solution (\ie, $SAT_{x^*}(M)$). To
determine if an iteration is a lucky iteration, we need the number of
violated constraints to be less than $1/3k$ of the cardinality of the
multi-set of constraints. We compute the sum total of the weights of
constraints that are satisfied by $x^*$ and the sum total of weights
of constraints not satisfied by $x^*$.

\begin{eqnarray*}
\sum_{v \in VIO_{x^*}(M)} v.w \leq \frac{1}{3k} \left(\sum_{v \in
    VIO_{x^*}(M)} v.w + \sum_{u \in SAT_{x^*}(M)} u.w\right)
\end{eqnarray*}
After rearranging the terms, we have
\begin{eqnarray*}
\sum_{v \in VIO_{x^*}(M)} v.w  \leq  \frac{1}{3k-1}{\sum_{u \in SAT_{x^*}(M)} u.w}
\end{eqnarray*}

Hence, if the sum of the weights of the violated constraints is less
than $\frac{1}{3k-1}$ of the sum of the weights of the satisfied
constraints, then it is a lucky iteration. Finally, the task of adding
the violated constraint again to the set $M$ is equivalent to doubling
the weights of the violated constraints.
Algorithm~\ref{alg:solvesample} presents our procedure for solving the
sample, identifying whether the iteration is a lucky iteration, and
updating the weights of the violated constraints.
Figure~\ref{fig:wrs} illustrates weighted random sampling and doubling
of the weights for the violated constraints with lucky iterations.

This entire process repeats until we find a polynomial that satisfies
all constraints (\ie, when the system is full-rank) or produces a
polynomial that violates at most a few points or exceeds the
user-specified threshold for the number of iterations. When the
algorithm exceeds the number of iterations without producing a
polynomial, we increment the number of terms used for the smaller
bitwidth representations in the progressive polynomial. We increase
the number of terms used for the largest representation when we are
unable to find a progressive polynomial after increasing the terms
used for the smaller representations.


When the system of linear equations is full-rank, then the above
procedure will find the unique polynomial. In many cases, the system
of equations may not be full-rank for a given number of terms in the
polynomial. Rather than increasing the number of terms, we also accept
a polynomial that satisfies all constraints except a few constraints
(\eg, typically 1-4 inputs in our experiments). For some elementary
functions, we also split the reduced inputs into two to four
sub-domains and generate polynomials for them to reduce the number of
terms.
In summary, we generate piecewise polynomials with fewer sub-domains
(\ie, 1 to 4) in comparison to our prior work in the \rlibm project
that generates numerous sub-domains (\eg, $2^{14}$ domains for $e^x$)
while also solving a much more challenging system of linear
inequalities with the progressive polynomial formulation.

\subsection{A Sketch of the Proof}
\label{proof}


The proof that our algorithm finds the solution for a system of
``full-rank'' linear constraints and terminates in $6k\log n$
iterations in expectation immediately follows from the proof of the
Clarkson's method~\cite{clarkson:vegas:jacm:1995}. We provide a sketch
of the proof for completeness. Here, $k$ is the number of terms in the
largest representation for the progressive polynomial. The proof
specifically relies on the following two lemmas.

\begin{lemma}{}
  \label{lemma:sample-solution}
There exist a set of $k$ constraints such that if we find an optimal
solution with respect to them, it will also be a feasible and optimal
solution for the entire set of $M$ constraints.
\end{lemma}

This lemma holds because the optimum value of a linear program is
always located on a vertex, which corresponds to $k$-strict
constraints.

\begin{lemma}{}
  \label{lemma:constraint-sampling}
Suppose we have a multi-set $M$ with $n$ constraints. If we sample
$6k^2$ constraints $S$ from $M$ and compute the optimum solution $x^*$
on $S$, then with probability at least $1/2$, $x^*$ can only violate
$1/3k$ constraints in $M$.
\end{lemma}

\textbf{Proof that the algorithm solves the system in $6k\log n$
  iterations in expectation.}
Let us consider the basis $B$ for the optimal solution in $M$, which
follows from Lemma~\ref{lemma:sample-solution}. Here, $B \subseteq
M$. Initially, $B$ has $k$ constraints as the rank of the system of
linear constraints is $k$ (\ie, $|B| = k$).  On every lucky iteration,
we double the constraints violated in $B$ (\ie $B$ is also a
multi-set). After $k$ lucky iterations, the number of constraints in
$B$ is at least $2k$. Similarly, the number of constraints in $B$ is
at least $2^2 \cdot k$ after $2k$ lucky iterations. Generalizing, the
number of constraints in $B$ is at least $2^{i} \cdot k$ after $k
\cdot i$ lucky iterations. Hence, $|B| \geq 2^{i} \cdot k \geq 2^i$.

From Lemma~\ref{lemma:constraint-sampling}, an iteration is lucky with
probability $1/2$, where the solution $x^*$ for the sample only
violates $1/3k$ or fewer constraints in the multi-set $M$. Hence, $M$
grows slowly. After $k\cdot i$ lucky iterations, size of the multi-set
$M$ is at most $\left(1+1/3k\right)^{k\cdot i} n$. From the Taylor's
series for $e^x$, we have $(1+x) \leq e^x$ for all $x$. Hence,
$(1+1/3k) \leq e^{1/3k}$.

\begin{eqnarray*}
  |M| \leq \left(1+1/3k\right)^{k \cdot i} n \leq \left(e^{1/3k}\right)^{k \cdot i} n \leq  e^{i/3} n
\end{eqnarray*}

The above two properties imply that there cannot be many lucky
iterations without finding a solution $x^*$ that satisfies all
constraints. In our setting, multi-set $B$ is a subset of multi-set
$M$. After $k\cdot i$ lucky iterations, the cardinalities of the sets
$B$ and $M$ should satisfy $2^i \leq n e^{i/3}$. When is $i \geq 3\log
n$, the above inequality is no longer true. Since we are exploring
$k\cdot i $ lucky iterations, the algorithm will terminate after
$3k\log n$ lucky iterations. Finally each iteration is lucky with
probability at least $1/2$ from
Lemma~\ref{lemma:constraint-sampling}. So, the algorithm terminates by
finding a solution that satisfies all constraints after $6k\log n$
iterations in expectation.


\textbf{Proof of Lemma~\ref{lemma:constraint-sampling}}.  To construct
the proof for this lemma, consider an artificial way of sampling as
follows: we first sample $r+1$ constraints $S'$ from $M$ and then
throw one of them out uniformly at random to get $S$. Here, $r$ is the
size of the sample $S$. This way of sampling $S'$ has the same
distribution as the original distribution of $S$. Let $X(S)$ be the
number of violated constraints when we sample $S$. For any constraint
$h\in M$, let $X(h,S)=1$ if and only if constraint $h$ is violated by
the optimum solution $x^*$ computed on $S$. Then, the expected value
of $X(S)$ is:

\begin{small}
\begin{eqnarray*}
E[X(S)] = \sum_{S} Prob(S)\sum_{h\notin S} X(h,S) =
\frac{1}{{|M| \choose r}}\sum_S\sum_{h\notin S} X(h,S)
\end{eqnarray*}
\end{small}

because the choice of $S$ is uniform over all $r$-subsets of $M$.
But interestingly, from our artificial way of sampling, we can also
write:

\begin{eqnarray*}
\sum_{S} \sum_{h\notin S}X(h,S) = \sum_{S'}\sum_{h\in S'} X(h,S'-h)
\end{eqnarray*}

Here, $S$ consists of all $r$-subsets of $M$ and $S'$ consists of all
$r+1$-subsets of $M$.

To understand when $X(h,S'-h) = 1$, fix a basis of $S'$ (as in
Lemma~\ref{lemma:sample-solution}). Then, $X(h,S'-h) = 1$ only when
$h$ belongs to this basis. But there are only $k$ choices of vectors
in this basis! So most of the time, the second summand is $0$. In
particular, we have:

\begin{eqnarray*}
E[X(S)] \leq \frac{1}{{|M| \choose r}} \sum_{S'} k
\end{eqnarray*}
Since the number of choices for $S'$ is $ \vert M \vert \choose r+1$,
so in total, we have:

\begin{eqnarray*}
E[X(S)] \leq \frac{{\vert M \vert \choose r+1}}{{ \vert M \vert \choose r}} k <  k\frac{\vert M\vert}{r+1}
\end{eqnarray*}

By Markov inequality, the probability that the value of $X(S)$ is at
least twice its expectation is at most $1/2$. Hence, we have:

\begin{eqnarray*}
Prob\left(X(S) > 2k\frac{\vert M\vert}{r+1}\right) < \frac{1}{2}
\end{eqnarray*}

Recall that we would like $X(S)$ to be at least $\vert M \vert/3k$. To
make $\frac{2k \vert M \vert}{r+1}< \frac{\vert M \vert}{3k}$, we can
pick $r=6k^2$, which is the size of the sample, so that the
probability of a lucky iteration is at least 1/2.

\section{Experimental Evaluation}
We describe the \tool prototype, experimental methodology, and the
results of our experiments to check both the correctness and
performance of our elementary functions.

\begin{table*}[!t]
  \small
  \caption{\normalsize Details of the polynomials generated by \tool
    in comparison to \rlibmall. For each function generated, we show the size
    of the piecewise polynomial, the maximum degree, and the number of
    terms (for bfloat16, tensorfloat32, and float types) in the
    polynomial. We also report the number of special case inputs to
    avoid increasing the degree of the polynomial approximation with \tool, and the size of the lookup
    tables for the coefficients of the generated polynomial approximations in bytes. We report the total
    reduction in memory for the lookup tables computed with \tool in comparison
    to \rlibmall.}
    \begin{tabular}[t]{| c | c | c | c | c |}
  \hline
  & \multicolumn{4}{| c |}{\rlibmall} \\
    \hline
      $f(x)$
      & \begin{tabular}{@{}c@{}}\# of \\ poly. \end{tabular}
      & \begin{tabular}{@{}c@{}}Deg-\\ree\end{tabular}
      & \begin{tabular}{@{}c@{}}\# of \\ terms \end{tabular}
      & \begin{tabular}{@{}c@{}c@{}}Poly. \\ mem. \\ use (B) \end{tabular}\\
      \hline
      $\mathbf{ln(x)}$ & $2^{10}$ & 3 & 3 & 24576\\ \hline
      $\mathbf{log_2(x)}$ & $2^{8}$ & 3 & 3 & 6144\\ \hline
      $\mathbf{log_{10}(x)}$ & $2^{8}$ & 3 & 3 & 6144\\ \hline
      $\mathbf{e^x}$ & $2^{8}$ & 4 & 5 & 10240\\ \hline
      $\mathbf{2^x}$ & $2^{8}$ & 3 & 4 &  8192\\ \hline
      $\mathbf{10^x}$ & $2^{9}$ & 3 & 4 & 16384\\ \hline
      $\mathbf{sinh(x)}$ & $2^6, 2^5$ & 5, 4 & 3, 3 & 2304\\ \hline
      $\mathbf{cosh(x)}$ & $2^6, 2^5$ & 5, 4 & 3, 3 & 2304\\  \hline
      $\mathbf{sinpi(x)}$ & $2^2, 2^2$ & 5, 4 & 3, 3 & 192\\ \hline
      $\mathbf{cospi(x)}$ & $2^2, 2^2$ & 5, 4 & 3, 3 & 192\\ \hline
   \end{tabular}
   \begin{tabular}[t]{| c | c | c | c | c | c | c | c | c | c |}
   \hline
   \multicolumn{10}{| c |}{\textbf{\tool}} \\
    \hline
      \multirow{3}{*}{\begin{tabular}{@{}c@{}}\# of \\ poly. \end{tabular}}
      & \multicolumn{3}{c |}{\multirow{2}{*}{Max. poly. degree}}
      & \multicolumn{3}{c |}{\multirow{2}{*}{\# of terms}}
      & \multirow{3}{*}{\begin{tabular}{@{}c@{}c@{}}\# of \\ special \\ inputs \end{tabular}}
      & \multirow{3}{*}{\begin{tabular}{@{}c@{}}Poly. \\ mem. \\ use (B) \end{tabular}}
      & \multirow{3}{*}{\begin{tabular}{@{}c@{}}\textbf{Reduction} \\ \textbf{in mem vs. }\\ \textbf{\rlibmall} \end{tabular}}\\
      & \multicolumn{3}{c |}{} & \multicolumn{3}{c |}{} & & & \\
      \cline{2-7}
      & \float & \tensorfloat & \bfloat & \float & \tensorfloat & \bfloat & & & \\ \hline
      $4$ & 5 & 5 & 0 & 5 & 5 & 0 & 13 & 360 & \textbf{68$\times$} \\ \hline
      $1$ & 5 & 3 & 0 & 5 & 3 & 0 & 0 & 40 & \textbf{154$\times$} \\ \hline
      $4$ & 6 & 3 & 0 & 6 & 3 & 0 & 3 & 216 & \textbf{28$\times$} \\ \hline
      $4$ & 6 & 4 & 3 & 7 & 5 & 4 & 0 & 160 & \textbf{64$\times$} \\ \hline
      $1$ & 6 & 3 & 2 & 7 & 4 & 3 & 0 & 56 & \textbf{146$\times$} \\ \hline
      $4$ & 6 & 4 & 3 & 7 & 5 & 4  & 1 & 176 & \textbf{93$\times$} \\ \hline
      $1, 1$ & 5, 4 & 3, 2 & 3, 2 & 3, 3 & 2, 2 & 2, 2 & 1, 1 & 80 & \textbf{29$\times$} \\ \hline
      $1, 1$ & 5, 4 & 3, 2 & 3, 2 & 3, 3 & 2, 2 & 2, 2 & 0, 0 & 48 & \textbf{48$\times$}  \\  \hline
      $1, 1$ & 5, 4 & 3, 2 & 3, 2 & 3, 3 & 2, 2 & 2, 2 & 0, 0 & 48 & \textbf{4$\times$} \\ \hline
      $1, 1$ & 5, 4 & 3, 2 & 3, 2 & 3, 3 & 2, 2 & 2, 2 & 0, 0 & 48 & \textbf{4$\times$} \\ \hline
   \end{tabular}
\label{tbl:statistics}
\end{table*}

 \textbf{Prototype.}  Our prototype, \tool, is a
progressive polynomial generator and a collection of correctly rounded
elementary functions. \tool contains multiple implementations for ten
elementary functions. A single progressive polynomial approximation
for each function produces the correctly rounded result for the 34-bit
FP representation that has 8-bits for the exponent with the
round-to-odd mode. It produces correctly rounded results for all FP
representations starting from 10-bits to 32-bits with all five
rounding modes in the IEEE standard. It also has progressive
performance with bfloat16 and tensorfloat32 types and produces
correctly rounded results for all inputs with them.  Correct and fast
polynomial approximations generated by \tool for $ln(x)$, $log_2(x)$,
and $log_{10}(x)$ are already part of LLVM's math
library~\cite{ly:logf:rlibm:2021,ly:log2f:rlibm:2021,ly:log10f:rlibm:2021}.

\tool uses the MPFR library~\cite{Fousse:toms:2007:mpfr} to compute
the oracle value of $f(x)$ for each representation. It uses an exact
rational LP solver, SoPlex~\cite{Gleixner:soplex:issac:2012}, to solve
constraints.  We use range reduction and output compensation functions
from our prior work in the \rlibm
project~\cite{lim:rlibm:popl:2021,lim:rlibm32:pldi:2021,lim:rlibmall:popl:2022}.
While evaluating the progressive polynomial, the bfloat16 and
tensorfloat32 inputs use only the first few terms of the progressive
polynomial. We perform polynomial evaluation, range reduction, and
output compensation using double precision. We use Horner's method to
evaluate polynomials~\cite{borwein:polynomials:book:1995}.
%

\textbf{Methodology.} We compare \tool's functions with
state-of-the-art libraries: Intel's double libm, glibc's double libm,
CR-LIBM~\cite{Daramy:crlibm:spie:2003}, and \rlibmall.  Intel's and
glibc's libm are mainstream libraries that are widely used for their
performance but do not provide correctly rounded results for all
inputs with any one rounding mode. CR-LIBM provides separate
implementations for each rounding mode for an elementary function that
produce the correctly rounded results for double precision.  It has
implementations for four out of the five rounding modes in the IEEE
standard and does not have an implementation for the
round-to-nearest-ties-to-away mode. \rlibmall produces correctly
rounded results for all $n$-bit FP representations and all five
rounding modes, where $10 \leq n \leq 32$.

We conducted our experiments on a 2.10GHz Intel Xeon Gold 6230R server
with 192GB of RAM running Ubuntu 20.04 that has both Intel turbo boost
and hyper-threading disabled to minimize perturbation.  We use the
publicly available CR-LIBM and \rlibmall versions.  We use Intel's
double libm from the oneAPI Toolkit and glibc's double libm from
glibc-2.31. The test harness for comparing glibc's libm, CR-LIBM, and
\rlibmall is built using the gcc-9.3.0 compiler with \texttt{-O0
  -frounding-math -fsignaling-nans} flags. The test harness for
comparing against Intel's libm is built using the \texttt{icc}
compiler with \texttt{-O0 -fp-model strict -no-ftz} flags because
Intel's libm is only available in the Intel's compiler. The
performance is measured using the number of cycles taken to compute
the result for each input using \texttt{rdtscp}. Then, we computed the
total time taken to compute the elementary functions for all inputs.

\textbf{Properties of \tool's polynomials.}
Table~\ref{tbl:statistics} provides details on the various properties
of the polynomial approximations generated by \tool in comparison to
\rlibmall.  With \tool, we tried to generate progressive polynomials
with the lowest degree with at most four sub-domains and with at most
four special case inputs per sub-domain (\ie, when the system is not
full-rank). We chose these thresholds because they can be implemented
efficiently with simple branches.  The range reduction strategy for
$sinh(x)$, $cosh(x)$, $sinpi(x)$, and $cospi(x)$ requires
approximations of two functions. We generated two polynomial
approximations for each elementary function.

\begin{table*}
	\footnotesize
	\caption{\small This table reports whether a library produces
          correctly rounded results for all inputs using \tool,
          glibc's double libm, Intel's double libm, CR-LIBM, and
          \rlibmall. Each sub-column also reports the ability to
          generate correctly rounded results for (1) bfloat16 and tensorfloat32
          results with the \RNE mode, (2) 32-bit float results with
          the \RNE mode, and (3) 32-bit float results with all five
          rounding modes. \cmark indicates that the library produces
          correctly rounded results for the given representation for all
          inputs. Otherwise, we use \xmark.}
        \vspace{-2mm}
\begin{tabular}{| c | c | c | c |}
	\hline
	$f(x)$ & \multicolumn{3}{| c |}{\textbf{\tool}} \\
	\hline
	\hline
	& \begin{tabular}{@{}c@{}}\bfloat  \& \\ \tensorfloat rn \end{tabular}  & 
	\begin{tabular}{@{}c@{}}\float \\ \RNE\end{tabular} & 
	\begin{tabular}{@{}c@{}}\float \\ all rm\end{tabular} \\
	\hline	
	$\mathbf{ln(x)}$ & \cmark & \cmark & \cmark \\ \hline
	$\mathbf{log_2(x)}$ & \cmark & \cmark & \cmark  \\ \hline
	$\mathbf{log_{10}(x)}$ & \cmark & \cmark & \cmark  \\ \hline
	$\mathbf{e^{x}}$ & \cmark & \cmark & \cmark \\ \hline
	$\mathbf{2^{x}}$ & \cmark & \cmark & \cmark  \\ \hline
	$\mathbf{10^{x}}$ & \cmark & \cmark & \cmark  \\ \hline
	$\mathbf{sinh(x)}$ & \cmark & \cmark & \cmark \\  \hline
	$\mathbf{cosh(x)}$ & \cmark & \cmark & \cmark  \\  \hline
	$\mathbf{sinpi(x)}$ & \cmark & \cmark & \cmark  \\ \hline
	$\mathbf{cospi(x)}$ & \cmark & \cmark & \cmark  \\ \hline
\end{tabular}
\begin{tabular}{| c | c | c |}
	\hline
	\multicolumn{3}{| c |}{glibc double libm} \\
	\hline
	\hline
	\begin{tabular}{@{}c@{}}\bfloat  \& \\ \tensorfloat rn \end{tabular}  & 
	\begin{tabular}{@{}c@{}}\float \\ \RNE\end{tabular} & 
	\begin{tabular}{@{}c@{}}\float \\ all rm\end{tabular} \\
	\hline
	\cmark & \xmark & \xmark \\ \hline
	\cmark & \cmark & \cmark  \\ \hline
	\cmark & \xmark & \xmark  \\ \hline
	\cmark & \cmark & \xmark \\ \hline
	\cmark & \xmark & \xmark  \\ \hline
	\cmark & \cmark & \xmark  \\ \hline
	\cmark & \xmark & \xmark \\  \hline
	\cmark & \cmark & \xmark  \\  \hline
	\NA & \NA & \NA  \\ \hline
	\NA & \NA & \NA  \\ \hline
\end{tabular}
\begin{tabular}{| c | c | c |}
	\hline
	\multicolumn{3}{| c |}{Intel double libm} \\
	\hline
	\hline
	\begin{tabular}{@{}c@{}}\bfloat  \& \\ \tensorfloat rn \end{tabular}  & 
	\begin{tabular}{@{}c@{}}\float \\ \RNE\end{tabular} & 
	\begin{tabular}{@{}c@{}}\float \\ all rm\end{tabular} \\
	\hline
	\cmark & \xmark & \xmark \\ \hline
	\cmark & \cmark & \cmark  \\ \hline
	\cmark & \xmark & \xmark  \\ \hline
	\cmark & \cmark & \xmark \\ \hline
	\cmark & \xmark & \xmark  \\ \hline
	\cmark & \cmark & \xmark  \\ \hline
	\cmark & \xmark & \xmark \\  \hline
	\cmark & \cmark & \xmark  \\  \hline
	\cmark & \cmark & \xmark  \\ \hline
	\cmark & \cmark & \xmark \\ \hline
\end{tabular}
\begin{tabular}{| c | c | c |}
	\hline
	\multicolumn{3}{| c |}{CR-LIBM} \\
	\hline
	\hline
	\begin{tabular}{@{}c@{}}\bfloat  \& \\ \tensorfloat rn \end{tabular}  & 
	\begin{tabular}{@{}c@{}}\float \\ \RNE\end{tabular} & 
	\begin{tabular}{@{}c@{}}\float \\ all rm\end{tabular} \\
	\hline
	\cmark & \xmark & \xmark \\ \hline
	\cmark & \cmark & \cmark  \\ \hline
	\cmark & \xmark & \xmark  \\ \hline
	\cmark & \cmark & \cmark\\ \hline
	\NA & \NA & \NA  \\ \hline
	\NA & \NA & \NA  \\ \hline
	\cmark & \xmark & \xmark \\  \hline
	\cmark & \cmark & \cmark  \\  \hline
	\cmark & \cmark & \xmark  \\ \hline
	\cmark & \cmark & \xmark  \\ \hline
\end{tabular}
\begin{tabular}{| c | c | c |}
	\hline
	\multicolumn{3}{| c |}{\rlibmall} \\
	\hline
	\hline
	\begin{tabular}{@{}c@{}}\bfloat  \& \\ \tensorfloat rn \end{tabular}  & 
	\begin{tabular}{@{}c@{}}\float \\ \RNE\end{tabular} & 
	\begin{tabular}{@{}c@{}}\float \\ all rm\end{tabular} \\
	\hline
	\cmark & \cmark & \cmark \\ \hline
	\cmark & \cmark & \cmark  \\ \hline
	\cmark & \cmark & \cmark  \\ \hline
	\cmark & \cmark & \cmark \\ \hline
	\cmark & \cmark & \cmark  \\ \hline
	\cmark & \cmark & \cmark  \\ \hline
	\cmark & \cmark & \cmark \\  \hline
	\cmark & \cmark & \cmark  \\  \hline
	\cmark & \cmark & \cmark  \\ \hline
	\cmark & \cmark & \cmark  \\ \hline
\end{tabular}

\label{tbl:accuracy}
\end{table*}

\textbf{Significant reduction in memory usage.} In contrast to
\rlibmall, \tool generates a single polynomial or a piecewise
polynomial with at most 4 sub-domains. \tool's polynomials require
only 123 bytes on average per function. In contrast, \rlibmall's
polynomials need 7667 bytes (7.5KB) on average per function. \tool's
polynomials reduce total storage needs by 62$\times$ on average compared to
\rlibmall.

\tool was able to generate a single progressive polynomial that
produces correctly rounded results without any special case inputs
for $log_2(x)$, $2^x$, $cosh(x)$, $sinpi(x)$, and $cospi(x)$, which
implies that the system is full-rank.  When we experimented with
\rlibmall's polynomial generation, it was not able to generate a
single polynomial for all functions except $log_2(x)$.  \tool
generates these progressive polynomials very quickly: only 19 minutes
on average per function. This shows the effectiveness of the \tool's
fast randomized algorithm for solving the set of constraints.

\textbf{Terms needed by bfloat16 and tensorfloat32.} When \tool
generates progressive polynomials, it indicates the number of terms
necessary to evaluate to produce correctly rounded results for the
bfloat16 and the tensorfloat32 types. Table~\ref{tbl:statistics} also
reports the number of terms that we need to evaluate in the
progressive polynomial to produce the correctly rounded bfloat16 and
the tensorfloat32 results. Surprisingly, a single term (first term) is
sufficient to produce correctly rounded results for all bfloat16
inputs with $ln(x)$, $log_2(x)$, and $log_{10}(x)$. In contrast,
\rlibmall's functions for $ln(x)$, $log_2(x)$, and $log_{10}(x)$ have
to evaluate a degree-3 polynomial to produce correctly rounded
bfloat16 results. The number of terms needed for bfloat16 and
tensorfloat32 are lower than the terms needed for computing correctly
rounded results for the 34-bit float with the round-to-odd mode except
where tensorfloat32 needs all terms for $ln(x)$.

\textbf{Does \tool produce correct results?}  Table~\ref{tbl:accuracy}
reports the summary of our evaluation to check whether \tool and other
existing libraries produce correctly rounded results for various
representations and rounding modes. All libraries produce correctly
rounded results for bfloat16 and tensorfloat32 results using the
round-to-nearest-ties-to-even (\RNE) mode. Glibc's double libm,
Intel's double libm, and CR-LIBM do not produce correctly rounded
results for 32-bit float inputs for several elementary functions and
various rounding modes. Even though CR-LIBM is a correctly rounded
library for double precision, it produces wrong results when it is
re-purposed for 32-bit floats due to double rounding errors. Both
\tool and \rlibmall produce correctly rounded float results for all
inputs and all standard rounding modes. More importantly, \tool is
able to produce correctly rounded bfloat16 and tensorfloat32 results
even when evaluating only the first few terms of the generated
progressive polynomial approximations.

\begin{figure*}
  \small
  \begin{subfigure}[b]{0.99\columnwidth}
    \caption{Speedup over glibc's double libm}
    \includegraphics[width=\linewidth]{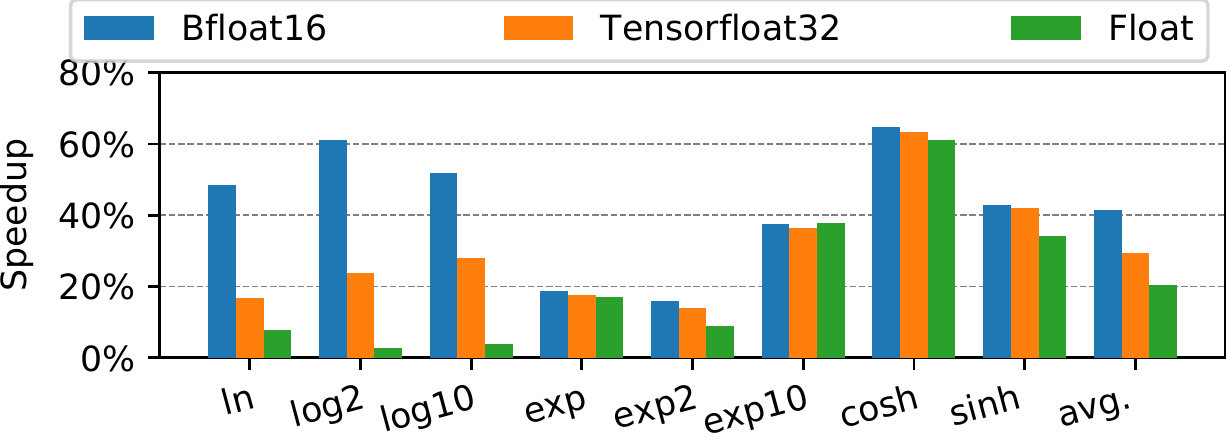}
  \end{subfigure}
  \begin{subfigure}[b]{0.99\columnwidth}
    \caption{Speedup over Intel's double libm}
    \includegraphics[width=\linewidth]{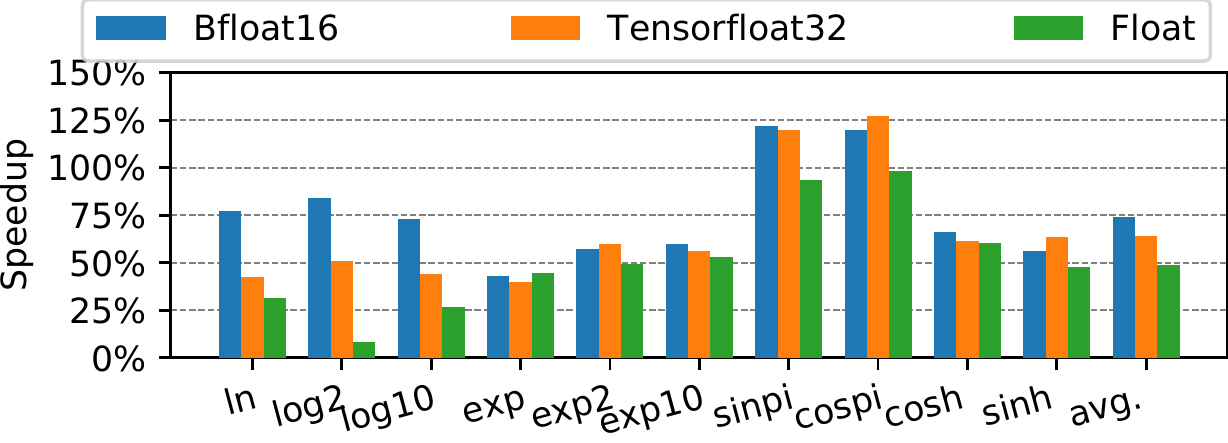}
  \end{subfigure}
  
  \vspace{1em}
  
  \begin{subfigure}[b]{0.99\columnwidth}
    \caption{Speedup over CR-LIBM}
    \includegraphics[width=\linewidth]{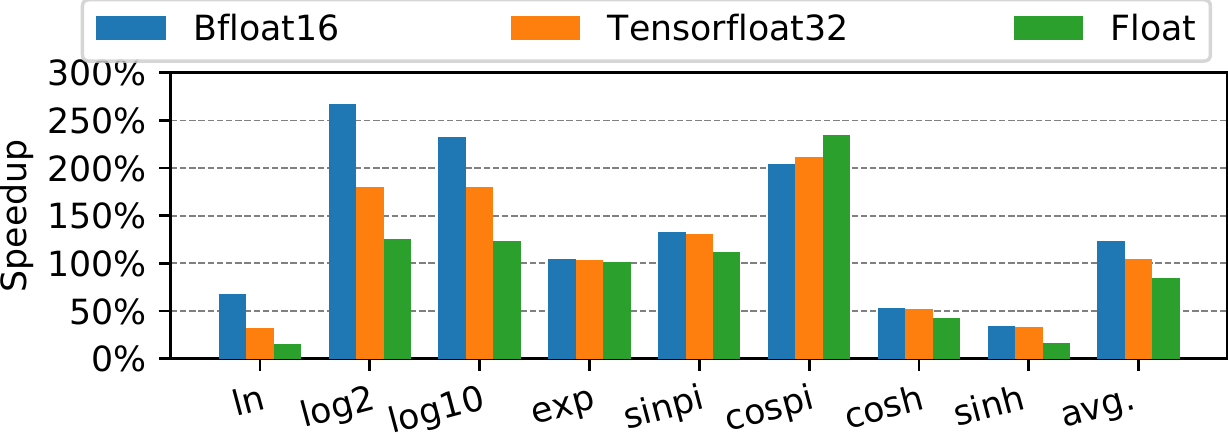}
  \end{subfigure}
  \begin{subfigure}[b]{0.99\columnwidth}
    \caption{Speedup over RLibm-all}
    \includegraphics[width=\linewidth]{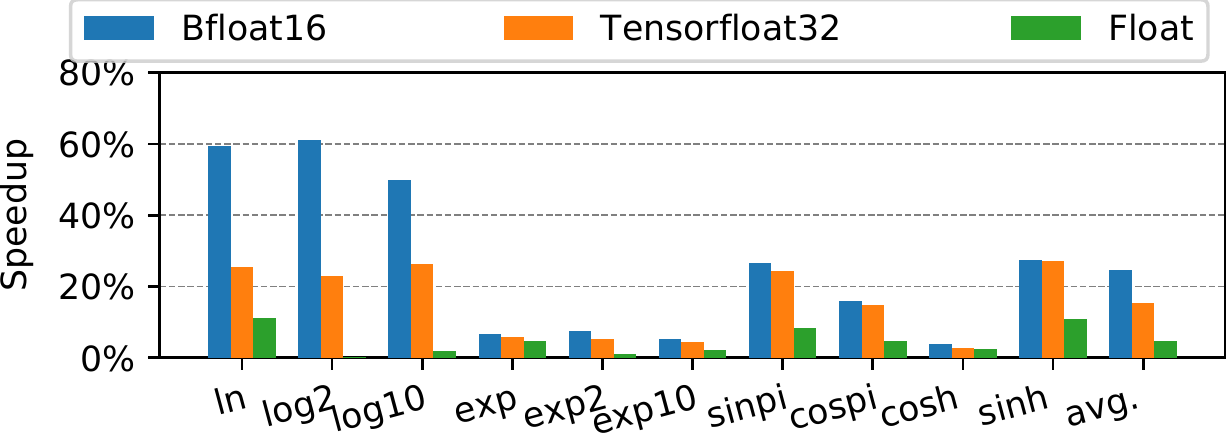}
  \end{subfigure}
  \caption{\small Computational speedup of \tool's progressive polynomial approximations in comparison
    to (a) glibc's double libm, (b) Intel's double libm, (c) CR-LIBM,
    and (d) \rlibmall. The left-most bar, middle bar, and the
    right-most bar in each cluster measures the speedup of \tool's
    bfloat16, tensorfloat32, and float elementary functions implemented as a
    progressive polynomial approximation.}
  \label{fig:prog_speedup_perc}
  \vspace{-2mm}
\end{figure*}

\textbf{Performance evaluation of \tool.}
Figure~\ref{fig:prog_speedup_perc} reports the speedup obtained with
\tool's functions when compared to various state-of-the-art
libraries. Figure~\ref{fig:prog_speedup_perc}(a) presents the speedup
of \tool's bfloat16 functions (left bar in each cluster),
tensorfloat32 functions (middle bar in each cluster), and float
functions (right bar in each cluster) over glibc's double libm. On
average, \tool's bfloat16, tensorfloat32, and float functions are
$42\%$, $29\%$, and $20\%$ faster over glibc's double library,
respectively. Similarly, Figure~\ref{fig:prog_speedup_perc}(b)
presents the speedup of \tool's functions over Intel's double library. On
average, \tool's bfloat16, tensorfloat32, and float functions are
$74\%$, $64\%$, and $49\%$ faster over Intel's double math
library. Intel's double library produces more accurate results
compared to glibc's double library and is slightly slower compared to
glibc's double library. Hence, \tool has more speedup over Intel's
double library compared to glibc's double library.

Figure~\ref{fig:prog_speedup_perc}(c) reports the speedup with \tool
when compared to CR-LIBM. On average, \tool's bfloat16, tensorfloat32,
and float functions are $123\%$, $105\%$, and $85\%$ faster over
CR-LIBM functions.

Figure~\ref{fig:prog_speedup_perc}(d) shows the speedup of \tool's
functions over \rlibmall. On average, \tool's bfloat16, tensorfloat32,
and float functions have $25\%$, $16\%$, and $5\%$ speedup over
\rlibmall. While \tool and \rlibmall's functions produce correctly
rounded results for all inputs, glibc's double libm, Intel's double
libm, and CR-LIBM are slower and do not produce correctly rounded results for
all inputs.

\tool generates significantly smaller piecewise polynomial approximations compared
to \rlibmall, which results in fewer memory accesses, producing
speedups even with the float functions. \rlibmall's $ln(x)$ function
has a piecewise polynomial of $2^{10}$ sub-domains whereas \tool's
$ln(x)$ function has a piecewise polynomial with $4$
sub-domains. Hence, \tool's float functions for $ln(x)$ are $11\%$
faster over \rlibmall. Similarly, \tool's $sinh(x)$ function uses two
single polynomials compared to \rlibmall's piecewise polynomials with
sizes of $2^{6} + 2^5$ (\ie, 96) sub-domains. Hence, \tool's $sinh(x)$
reports $11\%$ speedup over \rlibmall.

Even though the degree of the piecewise polynomials are smaller with
\rlibmall when compared to \tool for $e^x$ and $2^x$, \tool's
functions are $1\%$ and $2\%$ faster because the benefit from storing
fewer coefficients subsumes the overhead of evaluating a higher degree
polynomial.

\textbf{Progressive performance.}  Our performance evaluation
demonstrates that \tool's progressive polynomial approximations have better performance
for bfloat16 and tensorfloat32 types when compared to the float
type. \tool's bfloat16 functions show the highest speedup followed by
tensorfloat32, highlighting the progressive nature. \tool's $ln(x)$,
$log_2(x)$, and $log_{10}(x)$ functions for bfloat16 are $60\%$,
$61\%$, and $50\%$ faster over \rlibmall functions, respectively.
Although \rlibmall produces correctly rounded results for all bfloat16 inputs,
it requires evaluating the entire polynomial that results in some
performance loss.
In summary, \tool produces a single progressive polynomial
approximation that produces correctly rounded results for all inputs
with multiple representations and multiple rounding modes. Its float
functions are faster than state-of-the-art math
libraries. Furthermore, smaller representations are significantly
faster demonstrating progressive performance.

\section{Related Work}
Approximating and validating elementary functions is a well-studied
problem~\cite{Jeannerod:sqrt:tc:2011, Bui:exp:ccece:1999,
  Abraham:fastcorrect:toms:1991,
  Daramy:crlibm:spie:2003,Fousse:toms:2007:mpfr,Muller:elemfunc:book:2005,Trefethen:chebyshev:book:2012,
  Remes:algorithm:1934, Olga:metalibm:icms:2014,
  Brunie:metalibm:ca:2015, harrison:hollight:tphols:2009,
  Harrison:expproof:amst:1997, Harrison:verifywithHOL:tphol:1997,
  Sawada:verify:acl:2002,Lee:verify:popl:2018}, which has been
feasible because of advances in range
reduction~\cite{Tang:log:toms:1990,Tang:TableLookup:SCA:1991,Tang:exp:toms:1989,762822,
  Cody:book:1980, Boldo:reduction:toc:2009}.  A number of correctly
rounded math libraries have also been
developed~\cite{Abraham:fastcorrect:toms:1991,
  Daramy:crlibm:spie:2003, lim:rlibm:popl:2021,
  lim:rlibm32:pldi:2021}.  A detailed survey is available in Muller's
seminal book~\cite{Muller:elemfunc:book:2005}.
We restrict our discussion to the most closely related work. 

CR-LIBM~\cite{Daramy:crlibm:spie:2003, Lefevre:toward:tc:1998} is a
correctly rounded double library that provides implementations for a
subset of the rounding modes.  CR-LIBM relies on
Sollya~\cite{Chevillard:sollya:icms:2010} to generate near mini-max
polynomial approximations. CR-LIBM computes and proves the error bound
on the polynomial evaluation using interval
arithmetic~\cite{Chevillard:infnorm:qsic:2007,
  Chevillard:ub:tcs:2011}. Double rounding errors can cause wrong
results when the CR-LIBM's result is rounded to a 32-bit float.

This paper is closely related to our prior work in the \rlibm
project~\cite{lim:rlibm:popl:2021, lim:rlibm32:pldi:2021,
  lim:rlibmall:popl:2022, lim:rlibm:phdthesis:2021}. Like the prior
work in the \rlibm project, we also approximate the correctly rounded
result using an LP formulation. We also use \rlibm's range reduction
strategies. We use the idea of creating a single polynomial
approximation that produces correctly rounded results for multiple
representations and rounding modes from
\rlibmall~\cite{lim:rlibmall:popl:2022}.  We advance ideas from the
\rlibm project by generating faster polynomial approximations with a
novel method for solving linear constraints that provide progressive
performance with smaller bitwidth representations.

\section{Conclusion}
This paper proposes a novel type of polynomial approximations, termed
\emph{progressive polynomials}, that produce correctly rounded results
for multiple representations and rounding modes. An elegant property
of the progressive polynomial is that evaluating the first few terms
produces correctly rounded results for smaller representations. To
generate such progressive polynomials, we propose a fast algorithm for
polynomial generation that generates an order of magnitude smaller
lookup tables than the state-of-the-art method. \tool's polynomials
are faster than all mainstream and/or correctly rounded libraries.
We have already incorporated a few polynomial approximations from this
project in mainstream libraries. We believe this is the next logical
step in mandating correctly rounded elementary functions at least for
representations up to 32-bits.

\begin{acks}                            
  We thank Sepehr Assadi for his assistance with the proof of
  Clarkson's algorithm.  We thank John Gustafson for his inputs on the
  Minefield method and the posit representation.  We thank Fan Long
  and the PLDI reviewers for their feedback on a draft of this paper
  submitted to PLDI~\cite{aanjaneya:rlibm-prog:pldi:2022}.  This
  material is based upon work supported in part by the
  \grantsponsor{GS100000001}{National Science
    Foundation}{http://dx.doi.org/10.13039/100000001} under Grant
  No.~\grantnum{GS100000001}{1908798}, Grant
  No.~\grantnum{GS100000001}{2110861}, and Grant
  No.~\grantnum{GS100000001}{1917897}.
  Any opinions, findings, and conclusions or recommendations expressed
  in this material are those of the authors and do not necessarily
  reflect the views of the National Science Foundation.
\end{acks}

\bibliography{reference.bib}
\end{document}